# HCI Papers Cite HCI Papers, Increasingly So

Xiang 'Anthony' Chen
UCLA HCI Research
xac@ucla.edu

## ABSTRACT

To measure how HCI papers are cited across disciplinary boundaries, we collected a citation dataset of CHI, UIST, and CSCW papers published between 2010 and 2020. Our analysis indicates that HCI papers have been more and more likely to be cited by HCI papers rather than by non-HCI papers.

## CCS CONCEPTS

• **Human-centered computing**;

## KEYWORDS

HCI, Citation Metrics, Discipline, Impact

## 1 INTRODUCTION

Human-Computer Interaction (HCI) is often considered as one of the most interdisciplinary research fields. Thus we can expect that HCI papers should draw citation interests from other disciplines. Unfortunately, conventional citation metrics, such as h-index or i-10-index, only consider the number of citations but not the sources of citations. To address this limit, prior work measured HCI's impact on the industry by analyzing patents' citations of HCI papers [1]. In terms of how HCI impacts other academic fields, Wang *et al.* analyzed citation diversity by breaking down what fields cite and are cited by accesibility research [7]. While similar, our work differs from [7] in that we focus on surfacing the trend of HCI research drawing citations from non-HCI fields with comprehensive, multi-fold analyses.

Our goal is to measure the proportion of HCI papers' citations coming from outside of HCI venues. We first compiled a list of core HCI venues to define which fields are considered "non-HCI": although such a binary distinction oversimplifies the relationships amongst disciplines, it serves as a reasonable starting point to analyze how one field's citation influence reaches beyond its "comfort zone". We then collected citations data of papers from CHI, UIST, and CSCW (hereafter referred to as **HCI papers**) published between 2010 and 2020[1]. Our analysis found that—

- More recently-published HCI papers were cited less outside of the core HCI venues;
- Amongst all HCI papers cited in a given year, the earlier-published papers were cited more outside of the core HCI venues;
- In the more recent years, HCI papers seem less likely to be cited by papers outside of the core HCI venues.

## 2 DATASET

### 2.1 Core HCI Venues

We first need to define what is considered an HCI paper so that we can know which are HCI *vs.* non-HCI citations. To the best of our knowledge, there is no official "catalog" of all venues that publish HCI papers. Thus we refer to two sources to compile a list of core HCI venues: the SIGCHI-(co)sponsored conferences [5] and Google Scholar's top publications under the "Engineering & Computer Science - Human Computer Interaction" category [4].

Table 1 shows the compiled core HCI venues For each venue, we manually extract a subset of words from its official name to serve as a unique identifier (*e.g.*, "Engineering Interactive Computing Systems" for EICS), which we later use in keyword-matching to determine whether a citation is from one of these core venues.

Admittedly, what is not in this list is not necessarily "non-HCI"; it is just not included by the two sources we chose. Nonetheless, we argue that this list approximates the "boundary" between HCI and the non-HCI world. The more we see citations of an HCI paper coming outside of this list, the more likely this HCI paper influences research beyond HCI. We hereafter refer to papers published outside of this list as **non-HCI papers**.

### 2.2 Citations of CHI, UIST, and CSCW Papers

Next, we collected DOI data of CHI, UIST and CSCW papers between 2010 to 2020[2] from the "What The HCI" [6] website maintained by Kashyap Todi. We chose CHI, UIST and CSCW because they are three of the most well-recognized HCI venues. It is certainly possible to acquire DOI data of other HCI venues (*e.g.*, TOCHI) from other sources (*e.g.*, ACM Digital Library).

Then, we port the DOI data into the Citation Chaser app [2] developed by Neal Haddaway, which uses the Lens.org API [3] to retrieve citations of a paper based on its DOI. All the citation data was in the .ris format and collected in January 2023.

## 3 ANALYSES & FINDINGS

Going through each citation entry in our dataset, for a paper that cites an HCI paper, we parse the name of that paper's publishing venue using simple keyword-matching to determine whether that paper was published in one of the core HCI venues.

---

[1] Our dataset and source code can be found at https://github.com/hotnAny/x-index

[2] Except for CSCW 2020 papers, which the website did not provide.

| | |
|---|---|
| ASSETS | ACM SIGACCESS Conference on Computers and Accessibility |
| AutomotiveUI | International Conference on Automotive User Interfaces and Interactive Vehicular Applications |
| AVI | International Conference on Advanced Visual Interfaces |
| C&C | Creativity and Cognition |
| CHI | ACM CHI Conference on Human Factors in Computing Systems |
| CHI PLAY | The Annual Symposium on Computer-Human Interaction in Play |
| CI | Collective Intelligence |
| COMPASS | ACM SIGCAS/SIGCHI Conference on Computing and Sustainable Societies |
| CSCW | Computer Supported Cooperative Work |
| DIS | Designing Interactive Systems |
| EICS | ACM SIGCHI Symposium on Engineering Interactive Computing Systems |
| ETRA | Symposium on Eye Tracking Research and Applications |
| GI | Graphics Interface |
| GROUP | ACM International Conference on Supporting Group Work |
| HRI | ACM/IEEE International Conference on Human-Robot Interaction |
| ICMI | ACM International Conference on Multimodal Interaction |
| IDC | Interaction Design and Children |
| IJHCI | International Journal of Human-Computer Interaction |
| IJHCS | International Journal of Human-Computer Studies |
| IMWUT | Proceedings of the ACM on Interactive, Mobile, Wearable and Ubiquitous Technologies |
| IMX | ACM International Conference on Interactive Media Experiences |
| INTERACT | Human-Computer Interaction – INTERACT |
| ISS | Interactive Surfaces and Spaces |
| ISWC | International Symposium on Wearable Computers |
| IUI | International Conference on Intelligent User Interfaces |
| MobileHCI | International Conference on Mobile Human-Computer Interaction |
| NA | Behaviour and Information Technology |
| NA | International Journal of Interactive Mobile Technologies |
| NA | Universal Access in the Information Society |
| NordiCHI | Nordic Conference on Human-Computer Interaction |
| OzCHI | Australian Conference on Computer-Human Interaction |
| RecSys | ACM Conference on Recommender Systems |
| SMC | Transactions on Human-Machine Systems |
| SUI | International Conference on Spatial User Interaction |
| TAC | Transactions on Affective Computing |
| TACCESS | Transactions on Accessible Computing |
| TEI | International Conference on Tangible, Embedded, and Embodied Interaction |
| TOCHI | Transactions on Computer-Human Interaction |
| Ubicomp | ACM International Joint Conference on Pervasive and Ubiquitous Computing |
| UIST | ACM Symposium on User Interface Software and Technology |
| UMAP | ACM Conference on User Modeling, Adaptation and Personalization |
| VR | Virtual Reality and 3D User Interfaces |
| VRST | ACM Symposium on Virtual Reality Software and Technology |

**Table 1: A list of core HCI venues.**

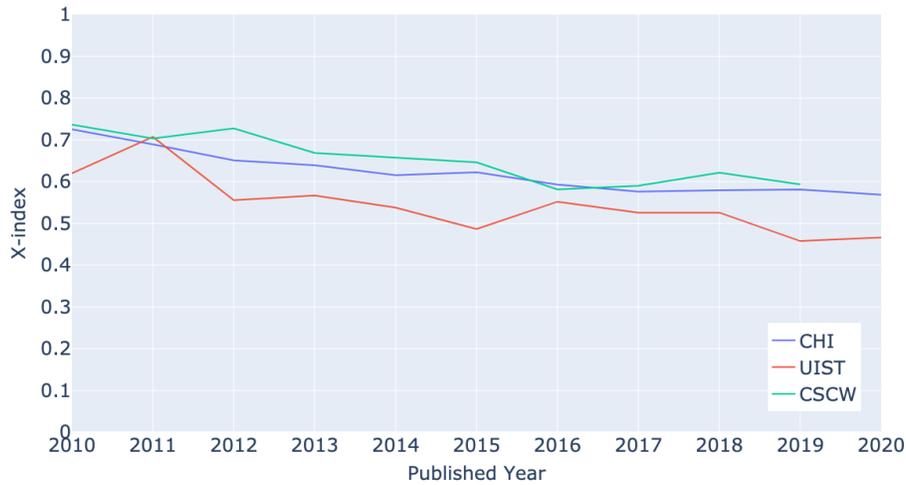

Figure 1: When counting all citations since publicaitons, more recently-published HCI papers were cited less outside of the core HCI venues.

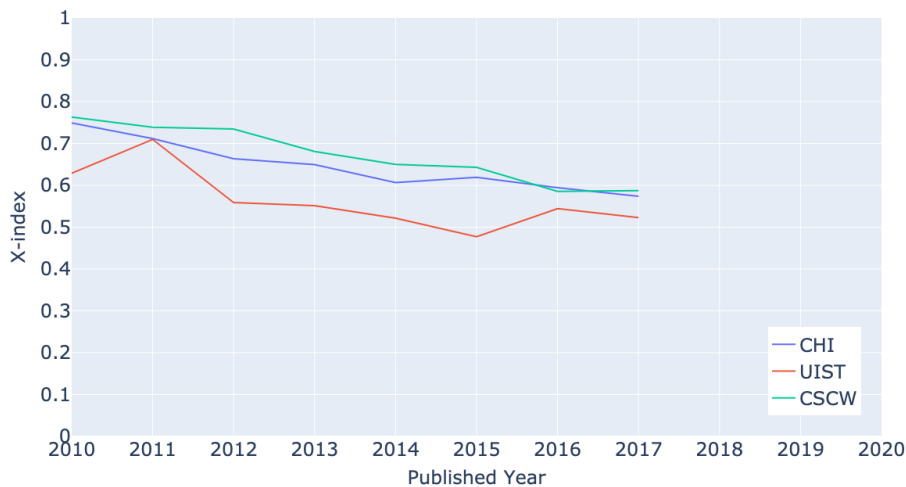

Figure 2: When counting only citations that occurred in the first five years after publicaitons, still more recently-published HCI papers were cited less outside of the core HCI venues.

### 3.1 More recently-published HCI papers were cited less outside of the core HCI venues

We analyzed each year's published HCI papers (*e.g.*, 11 years of UIST papers published between 2010 and 2020), as shown in Figure 1. Each point in Figure 1 is the non-HCI-over-HCI ratio based on citations of that year's published HCI papers up to when the data was collected (January 2023). We can see that all three HCI venues' papers have an decreasing ratio of getting cited by non-HCI paeprs over the years.

One might question that, since citation count is also a function of time, the more recently-published HCI papers might be disadvantaged simply because they have not existed long enough to attract citations. To resolve this doubt, we further consider citations that occurred only within the first five years after an HCI paper was published. For example, for CHI 2015 papers, we only consider citations of them that occurred between 2016 and 2020—in this way, CHI 2010 papers have no time advantage over CHI 2015 papers. This constraint narrows the overall analysis down to only HCI papers published from 2010 to 2017 because, later than that, our data (collected in January 2023) has less than five years of citation information. As shown in Figure 2, results indicate that, even accounting for how long an HCI paper has been around, the overall trend remains similar over the years.

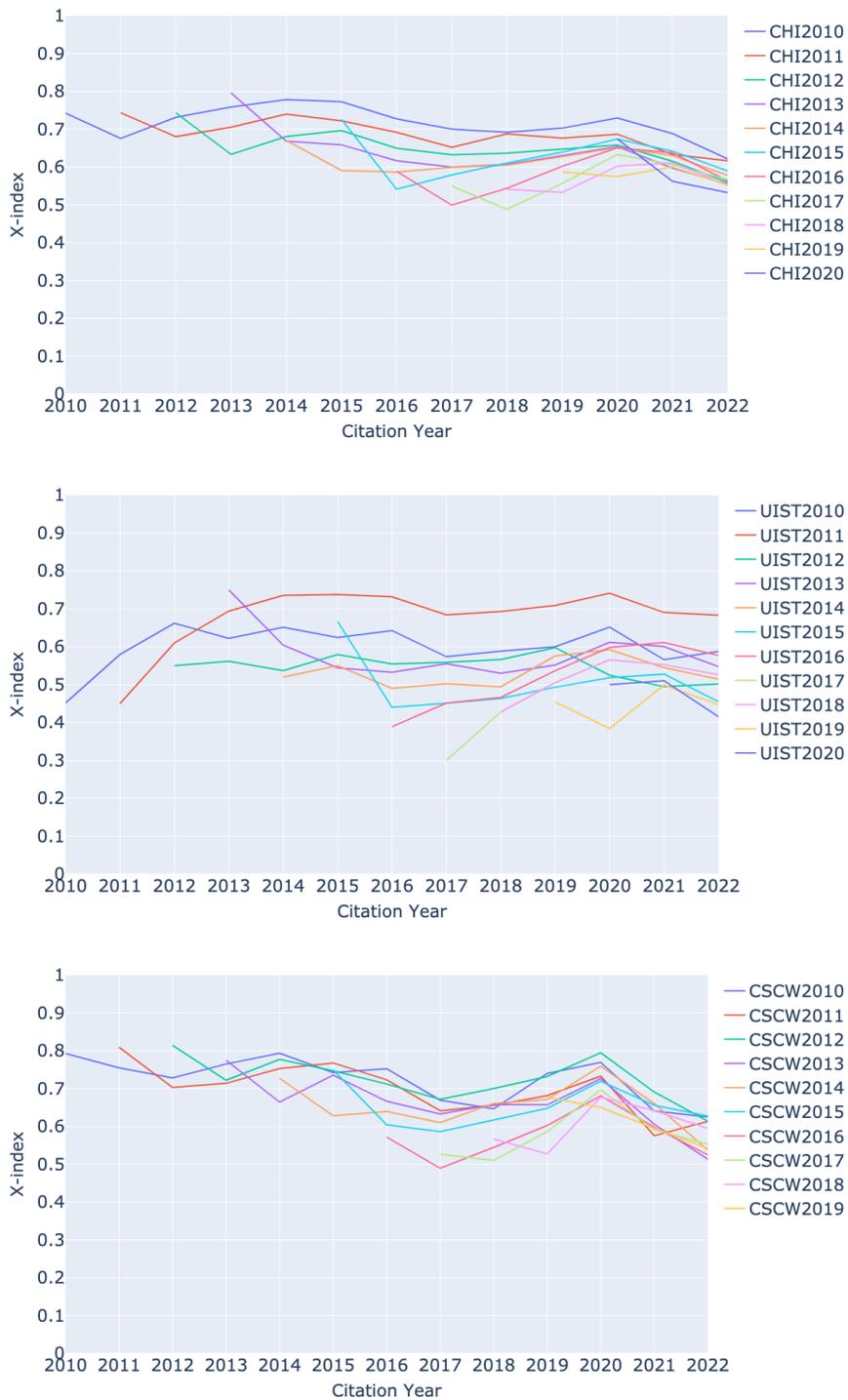

Figure 3: Amongst all HCI papers cited in a given year, the earlier-published papers were cited more outside of the core HCI venues.

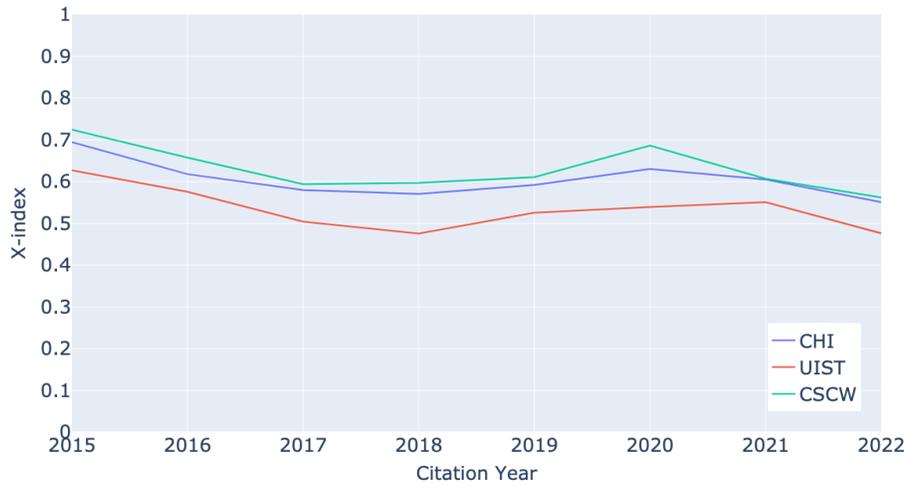

Figure 4: In the more recent years, HCI papers seem less likely to be cited by papers outside of our core HCI venue list.

## 3.2 Amongst all HCI papers cited in a given year, the earlier-published papers were cited more outside of the core HCI venues

To expand our analysis along the time dimension, we plot each year's published HCI papers' citations by non-HCI venues over the subsequent years, as shown in Figure 3.

Note that the $x$-axis in Figure 3 has changed: it is now citation year, not published year. For example, when some papers in 2019 cited a CSCW 2013 paper, the citation year is 2019 whereas 2013 is the publication year. The results show that, for most years' HCI papers, the ratio of non-HCI citations tends to flatten or slightly decrease over time, except for a few 'local' cases, *e.g.*, UIST 2011 had an increasing trend until 2014.

Further, amongst HCI papers cited in a given citation year, the earlier papers tend to have attracted more non-HCI citations than the later ones. Consider the citation year 2020. For CHI, the top-3 in 2020 are papers published in 2010, 2011, and 2015 whereas the bottom-3 are 2019, 2018, and 2017; For UIST, the top-3 in 2020 are papers published in 2011, 2010, and 2013 whereas the bottom-3 are 2019, 2020, and 2012; For CSCW, the top-3 in 2020 are papers published in 2012, 2010, and 2014 whereas the bottom-3 are 2019, 2018, and 2016.

## 3.3 In the more recent years, HCI papers seem less likely to be cited by papers outside of the core HCI venues

The previous analysis suggests that we can aggregate each year's citations of HCI papers so that we can see how the non-HCI papers' interest in citing HCI papers changed over the years. For each year, we consider that year's citations of HCI papers published in previous five years. Since we do not have citation information before 2010, our analysis can only begin from 2015 (A 2014 paper might cite a CHI 2009 paper but we do not have CHI 2009's citation information in our dataset). Figure 4 shows the results, which also exhibit a decreasing pattern over the years. In other words, in the more recent years, HCI papers seem less likely to be cited by papers outside of our core HCI venue list.

## 4 DISCUSSIONS
### 4.1 Limitations of the Analyses

Our classification of non-HCI papers is "impure" because it is based on a non-exhaustive list of core HCI venues. Thus some citations we count as non-HCI might actually have come from HCI venues not included in our list. A common example of this issue is various HCI workshops over the years. As a consequence, the resultant non-HCI citations ratio in our analyses is actually inflated—the real numbers should be lower if we could remove all the "noisy" HCI venues from the currently-considered non-HCI citations.

Second, we rely on Lens.org API [3] that does not seem to retrieve all citations exhaustively. We do not believe such a limitation invalidates our analyses unless this API acts biasedly, *i.e.*, intentionally misses only HCI citations. Verifying whether such biases exist would require "groundtruth" citation retrievals, which was unavailable at the time of this work. To overcome this limitation, future work can union multiple tools and sources (*e.g.*, Google Scholar, Semantic Scholar, and Microsoft Academic) to approximate a full coverage of citations.

Third, the ratio of citations by non-HCI venues is just a number and it would be great to perform more detailed analyses, such as breaking down the sources that cite CHI papers. Future work can employ a more intelligent method than our simple keyword-matching approach to dissect the disciplinarity of citation sources (*e.g.*, $X$% HCI, $Y$% Computer Vision, and $Z$% Psychology).

### 4.2 Interpreting the Results

Does our analyses mean that HCI has a decreasing impact across the disciplinary boundary? It seems so, at least within the academic world. Our analyses show that relatively fewer and fewer papers outside of the core HCI venues are citing HCI papers.

One counter-argument against the above could be that HCI venues themselves might have become more and more interdisciplinary. Thus citations coming from within the more recent HCI venues can also indicate impact across what used to be the disciplinary boundary.

It is also possible that publications in HCI venues are growing much faster than those in non-HCI venues. In that case, even if more and more non-HCI venues are citing HCI papers, the overall ratio would still decrease if the denominator (HCI + non-HCI citations) is increasing faster than the nominator (non-HCI citations).

It will be interesting for future work to perform the same analysis on other research fields. For example, according to Google Scholar, as of March 2023, CVPR has the highest h-index amongst all "Engineering & Computer Science" publications—Should we expect citations of CVPR papers to come from the broad enginering and computer science fields?

## 5 CONCLUSION

Curious to find out how much a field's papers are cited by papers outside of that field, our analyses of CHI, UIST, and CSCW papers (2010–2020) show that **HCI papers cite HCI papers, increasingly so**. While it seems reasonably acceptable for papers in one field to cite each other, HCI as an interdisciplinary field should aim beyond that. We hope our analyses can promote more awareness of HCI's impact beyond the disciplinary boundaries rather than remaining complacent with validating each other's work internally via having more and more HCI papers cite other HCI papers.

## REFERENCES
[1] Hancheng Cao, Yujie Lu, Yuting Deng, Daniel A. McFarland, and Michael S. Bernstein. 2023. Breaking Out of the Ivory Tower: A Large-scale Analysis of Patent Citations to HCI Research. https://doi.org/10.1145/3544548.3581108 arXiv:2301.13431 [cs].
[2] Neal Haddaway. 2023. citationchaser. https://estech.shinyapps.io/citationchaser/. (Accessed on 02/24/2023).
[3] Lens.org. 2023. The Lens - Free & Open Patent and Scholarly Search. https://www.lens.org/. (Accessed on 02/24/2023).
[4] Google Scholar. 2023. Human Computer Interaction - Google Scholar Metrics. https://scholar.google.com/citations?view_op=top_venues&hl=en&vq=eng_humancomputerinteraction. (Accessed on 02/24/2023).
[5] SIGCHI. 2023. Upcoming Conferences – ACM SIGCHI. https://sigchi.org/conferences/upcoming-conferences/. (Accessed on 02/24/2023).
[6] Kashyap Todi. 2023. WhatTheHCI? https://whatthehci.com/. (Accessed on 02/24/2023).
[7] Lucy Lu Wang, Kelly Mack, Emma J McDonnell, Dhruv Jain, Leah Findlater, and Jon E. Froehlich. 2021. A bibliometric analysis of citation diversity in accessibility and HCI research. In *Extended Abstracts of the 2021 CHI Conference on Human Factors in Computing Systems (CHI EA '21)*. Association for Computing Machinery, New York, NY, USA, 1–7. https://doi.org/10.1145/3411763.3451618